**Tunable low-phase-noise terahertz synthesis using a fiber-interferometer-referenced optical frequency comb and an EDFA-integrated Vernier cavity**

Masayuki Higaki[1], Yoshiaki Nakajima[2], Yu Tokizane[3,4], Kwangyun Jung[5], Ryosuke Nakamura[5], Naoya Kuse[3,4], and Takeshi Yasui[3,4]

[1]Graduate School of Sciences and Technology for Innovation, Tokushima University, 2-1, Minami-Josanjima, Tokushima, Tokushima 770-8506, Japan

[2]Department of Physics, Toho University, 2-2-1 Miyama, Funabashi, Chiba 274-8510, Japan

[3]Institute of Post-LED Photonics (pLED), Tokushima University, 2-1, Minami-Josanjima, Tokushima, Tokushima 770-8506, Japan

[4]Institute of Photonics and Human Health Frontier (IPHF), Tokushima University, 2-1, Minami-Josanjima, Tokushima, Tokushima 770-8506, Japan

[5]Sevensix Inc., 17F Roppongi Hills Mori Tower, 6-10-1 Roppongi, Minato-ku, Tokyo 106-6117, Japan

## ABSTRACT

We demonstrate a frequency-tunable, low-phase-noise terahertz (THz) synthesizer based on a fiber-interferometer-referenced optical frequency comb (OFC), an erbium-doped fiber amplifier (EDFA)-integrated Vernier cavity, and dual optical injection locking (OIL). The Vernier cavity converts the densely spaced modes of a fiber OFC with a repetition frequency of approximately 100 MHz into a filtered comb with gigahertz-scale effective mode spacing, facilitating selective single-mode OIL. Intracavity and extracavity EDFAs in the Vernier cavity enable a side-mode suppression ratio exceeding 70 dB. Independent control of the Vernier-cavity free spectral range and the OFC repetition frequency provides flexible Vernier filtering conditions for selecting master modes for dual OIL. Two selected comb modes are transferred to injection-locked optical carriers and photomixed in a uni-traveling-carrier photodiode to generate continuous-wave THz signals. Frequency-tunable THz generation is demonstrated at 230, 267, and 320 GHz. At 267 GHz, a signal-to-noise ratio exceeding 70 dB and a measured phase noise of −95.7 dBc/Hz at a 10-kHz offset are obtained; the phase-noise measurement is limited by the local-oscillator system. The proposed architecture provides a fiber-based approach to frequency-tunable, low-phase-noise THz synthesis by combining an optical-reference-based OFC with flexible Vernier-assisted comb-mode selection.

## 1. INTRODUCTION

Terahertz (THz) waves have attracted increasing attention for a wide range of applications, including wireless communications [1,2], spectroscopy [3,4], imaging [5,6], and radar systems [7,8]. As these applications advance, there is a growing demand for high-performance THz frequency synthesizers that combine high frequency stability, low phase noise, frequency traceability, and frequency tunability. In wireless communications, precise control of carrier frequencies is essential for efficient use of limited spectral resources and standardized frequency allocation, while in THz frequency-domain spectroscopy, accurate and traceable frequency references are required for reliable identification of spectral features. Conventional electronic approaches based on frequency multiplication of microwave sources provide frequency traceability but suffer from increasing phase noise as the multiplication factor increases, particularly in the THz frequency range. In contrast, photonic approaches offer favorable scalability toward higher frequencies while maintaining low phase noise, making them a promising platform for high-performance THz synthesis.

Among photonic approaches, optical frequency combs (OFCs) [9,10] provide a coherent bridge between the optical and RF/microwave frequency domains. When an OFC is stabilized to a microwave or RF reference, THz frequencies are generated through high-order frequency multiplication of the comb repetition rate ($f_{rep}$), resulting in a phase-noise penalty that increases with the multiplication factor. In contrast, an OFC stabilized to a low-noise optical reference can operate as an optical frequency divider (OFD), coherently transferring the superior frequency stability and low phase noise of the optical reference to the millimeter-wave and THz domains, with the phase-noise power reduced in proportion to the square of the corresponding optical-to-carrier frequency division factor [11,12]. By selecting two optical modes from such an optically referenced OFC and photomixing them in a photodetector, a continuous-wave (CW) THz signal corresponding to their frequency difference can be synthesized while preserving the low-noise characteristics of the optical reference. In particular, fiber-based OFCs with repetition rates of approximately 100 MHz provide densely spaced optical modes, offering high flexibility in selecting the frequency difference and thus the synthesized THz frequency. Therefore, optical-reference-based OFD using a fiber OFC provides a promising route toward low-phase-noise and frequency-tunable THz synthesis.

Optical-reference-based OFD has enabled exceptionally stable THz synthesis, but its practical implementation remains constrained by sophisticated optical-reference and comb-mode-selection systems. Shin *et al.* demonstrated an ultrastable THz synthesizer based on a fiber OFC stabilized to a high-finesse ultra-low-expansion (ULE) optical cavity, achieving THz synthesis over 0.10–1.10 THz with a fractional frequency instability of $4.4\times10^{-15}$ at 1 s at 0.66 THz [12]. In their system, individual modes of the densely spaced fiber OFC were isolated using narrowband filtering based on a combination of a fiber Fabry–Perot (FFP) cavity and a fiber Bragg grating (FBG), followed by optical injection locking (OIL) to distributed-feedback lasers for amplification and photomixing. While this approach provides outstanding frequency stability and spectral purity, the use of a high-finesse ULE cavity as the optical reference and narrowband filtering for individual comb-mode selection increases the complexity of the overall system. Therefore, a more practical architecture that preserves the low-noise advantages of optical-reference-based OFD while simplifying both the optical-reference and comb-mode-selection systems is highly desirable.

To address these limitations, we employ a long-fiber interferometer [13] as a practical fiber-based optical frequency reference and introduce an erbium-doped fiber amplifier (EDFA)-integrated Vernier cavity for flexible comb-mode selection. Vernier mode filtering effectively increases the mode spacing of a dense fiber OFC from approximately 100 MHz to the gigahertz range by suppressing intermediate comb modes, thereby facilitating stable single-mode OIL. Nakajima *et al.* developed an all-fiber mode-filtering technique [14] and subsequently demonstrated a phase-stabilized gigahertz-spaced frequency comb with a high side-mode suppression ratio (SMSR) [15]. In the present work, we extend this approach for THz synthesis by incorporating EDFAs and a variable optical delay line (ODL) into the Vernier cavity. The combination of intracavity and extracavity EDFAs compensates for optical losses associated with the filtering and recirculation processes and enhances mode suppression. The ODL provides control of the cavity free spectral range (FSR), while the repetition frequency $f_{rep}$ of the fiber OFC is independently controlled. Because the Vernier filtering condition is determined by the combination of the cavity FSR and $f_{rep}$, these two independently controllable parameters enable gigahertz-scale control of the filtering interval. This architecture combines high-SMSR mode filtering with flexible mode selection for subsequent OIL and tunable THz synthesis.

In this study, we demonstrate tunable low-phase-noise THz synthesis by combining a fiber-interferometer-referenced OFC with an EDFA-integrated Vernier cavity and dual OIL. High-SMSR Vernier mode filtering enables selective single-mode

OIL of two desired comb modes, and the resulting injection-locked optical carriers are photomixed in a uni-traveling-carrier photodiode (UTC-PD) to generate CW THz signals. THz generation is demonstrated at 230, 267, and 320 GHz, with a signal-to-noise ratio (SNR) exceeding 70 dB at 267 GHz. At 267 GHz, a measured phase noise of −95.7 dBc/Hz at a 10-kHz offset is obtained, although the measurement is limited by the phase noise of the local-oscillator system. These results demonstrate a fiber-based architecture for frequency-tunable, low-phase-noise THz synthesis using flexible Vernier-based comb-mode selection.

## 2. Experimental system and characterization

### 2.1 Operating principle of the tunable THz synthesizer

Figure 1 illustrates the overall configuration and operating principle of the proposed tunable low-phase-noise THz synthesizer. The system consists of a fiber-interferometer-stabilized optical reference, a fiber OFC, an EDFA-integrated tunable Vernier cavity, dual OIL lasers, and a UTC-PD. A narrow-linewidth CW laser stabilized to a long-fiber interferometer serves as a common optical frequency reference for both the fiber OFC and the Vernier cavity. The carrier-envelope-offset frequency ($f_{ceo}$) of the fiber OFC is independently locked to a rubidium (Rb) frequency reference. The fiber OFC is then phase-locked to the stabilized CW laser by controlling its $f_{rep}$, thereby transferring the frequency stability and low-phase-noise characteristics of the optical reference to the comb modes. In parallel, the Vernier cavity is stabilized to the same optical reference by Pound–Drever–Hall (PDH) locking. Referencing both the OFC and the Vernier cavity to the same optical reference maintains the relative frequency relationship between the OFC modes and the Vernier-cavity resonances, enabling stable Vernier mode filtering.

The mode-locked fiber OFC provides densely spaced optical modes with a repetition frequency $f_{rep}$ of approximately 100 MHz. When the fiber OFC is transmitted through the Vernier cavity, whose FSR differs from $f_{rep}$, intermediate comb modes are suppressed through the Vernier effect, resulting in a filtered comb with a mode spacing in the gigahertz range. The Vernier filtering facilitates stable single-mode OIL by strongly suppressing the optical power of unwanted comb modes adjacent to the selected master mode, thereby increasing the spectral contrast of the master signal. The intracavity and extracavity EDFAs compensate for cavity loss and enhances side-mode suppression. The cavity FSR can be adjusted using a variable ODL incorporated into the cavity, while $f_{rep}$ of the fiber OFC is independently controlled. The Vernier filtering condition is therefore determined by the combination of these two frequency parameters, providing flexibility in setting the effective mode spacing of the filtered comb for subsequent OIL.

Two different modes of the Vernier-filtered comb are subsequently used as master signals for two OIL lasers. Each selected comb mode injection-locks a slave laser, generating a high-power narrow-linewidth optical carrier that inherits the frequency and phase characteristics of the corresponding comb mode. The resulting two optical carriers, with optical frequencies $v_1$ and $v_2$, are selected with a frequency separation of few-hundred-gigahertz range and are simultaneously incident on the UTC-PD. Photomixing of the two carriers generates a continuous-wave THz signal at their difference frequency,

$$f_{THz} = \Delta\nu = |\nu_2 - \nu_1|. \quad (1)$$

Because the frequency stability and low-phase-noise characteristics originate from the common fiber-interferometer-stabilized optical reference and are transferred through the OFC to the two selected optical carriers, the generated THz signal reflects their relative frequency and phase characteristics. The following subsections characterize each key stage of this architecture, beginning with the fiber-interferometer-stabilized optical reference.

### 2.2 Fiber-interferometer-stabilized optical reference

Figure 2(a) shows the configuration of the fiber-interferometer-stabilized optical reference used as the common frequency reference in the proposed THz synthesizer. Two highly unbalanced fiber Michelson interferometers with the same fiber-delay lengths are employed in a cascaded stabilization configuration. In each interferometer, one arm is terminated by a Faraday rotator mirror (FRM), while the other arm contains an acousto-optic modulator (AOM) for optical-frequency shifting and is terminated by a second FRM. The heterodyne beat signal between the optical fields returning from the two arms is used to derive the error signal for laser-frequency stabilization. To suppress differential environmental perturbations, the two interferometers are housed in the same enclosure with vibration-mitigation measures and active temperature control. Under ambient temperature fluctuations of approximately 0.3 °C, the temperature variation inside the enclosure is maintained within 0.06 °C.

The two interferometers provide frequency -discrimination and are used for higher noise suppression over wide frequency range. As the laser current modulation bandwidth is limited to 30 kHz, AOM-based frequency feedback with 300 kHz modulation bandwidth was additionally used. Of course, only one interferometer error signal can drive both laser current and AOM frequency by dividing the feedback frequency band. However, enough feedback gain cannot be obtained not only at near frequency overlap band, but also at low frequency band by the low laser modulation bandwidth. By using two interferometers, the two feedback loops can be split completely, and sufficient feedback gain is obtained over wide frequency range. In the first stabilization stage, the frequency fluctuations of the CW laser are detected by fiber interferometer #1. The resulting error signal is fed back to the laser injection current, providing a relatively wide frequency-tuning range with low feedback gain and bandwidth of approximately 30 kHz. In the second stage, the residual frequency fluctuations are detected by fiber interferometer #2 with a fiber delay length of 70 m. The resulting error signal is fed back to an AOM for optical-frequency control, providing a relatively narrow tuning range with higher feedback gain and bandwidth of approximately 300 kHz. Thus, the two stabilization stages combine the wide tuning range of the injection-current feedback with the high feedback gain and bandwidth of the AOM-based control. Note that, 70 m fiber length was selected by compromising with frequency noise detection sensitivity and achievable maximum feedback bandwidth. The longer fiber length gives higher sensitivity, but the feedback bandwidth gets lower. Inverse of the round trip time for the 70 m fiber is 1.4 MHz and AOM feedback bandwidth is near 300 kHz. Therefore, this is the maximum length without affecting feedback bandwidth limited by the AOM. The fiber length of the first stage should be the same with the second stage. The length difference results in temperature frequency drift between the two stages. The limited-AOM tuning range cannot compensate such a large drift.

Figure 2(b) shows the single-sideband (SSB) phase-noise spectrum of the fiber-interferometer-stabilized optical reference. At an optical carrier frequency of approximately 194 THz, the stabilized CW laser exhibited an SSB phase noise of −83.6 dBc/Hz at an offset frequency of 10 kHz. This low-phase-noise optical signal serves as the common optical frequency reference for both the fiber OFC and the Vernier cavity, as illustrated in Fig. 1. Specifically, it is used for phase locking of the fiber OFC through control of $f_{rep}$ and for PDH locking of the Vernier cavity.

### 2.3 EDFA-integrated Vernier mode filtering

Figure 3(a) illustrates the principle of Vernier mode filtering. The fiber OFC has densely spaced modes with a repetition frequency $f_{rep}$ of approximately 100 MHz. By transmitting the OFC through a Vernier cavity whose FSR differs from $f_{rep}$, the OFC modes and cavity resonances periodically coincide owing to the Vernier effect [14,15]. The resulting Vernier filtering interval $f_{Vernier}$ is expressed as

$$f_{Vernier} = M \cdot f_{rep} = N \cdot FSR. \quad (2)$$

where *M* and *N* are coprime positive integers. Under this condition, every *M*-th OFC mode coincides with every *N*-th cavity resonance, while the intermediate OFC modes are suppressed. Thus, the Vernier filtering interval is determined by the relative relationship between the cavity FSR and the OFC repetition frequency rather than by either frequency alone. The resulting filtered comb has an effective mode spacing in the gigahertz range, which increases the spectral contrast of the selected comb modes and facilitates their use as master signals for single-mode OIL.

Figure 3(b) shows the experimental configuration of the EDFA-integrated Vernier cavity. The Vernier cavity is constructed using polarization-maintaining fiber (PMF; corning, Panda PM 1550), allowing the OFC and optical reference to propagate along the orthogonal polarization axes of the same fiber cavity. The fiber OFC is polarized along the slow axis of the PMF and coupled into the Vernier cavity, which incorporates a home-built intracavity EDFA, a variable ODL (Alnair, MDL-15-068-PM-1-FA-AG009; delay range = 0 – 680 ps), a piezoelectric transducer (PZT; Thorlabs, PC4QR; displacement range = 20.0 μm) for cavity-length control, 99:1 fiber couplers ((AFR, PMC-2-1550-01-P-2-L-Q-6, PMC-1-1550-01-P-2-L-Q-6), and polarization-control components, including a polarization beam splitter (PBS; AFR, PBS-55-P-2-L-3-Q) and polarizers (Ps; AFR, ILP-55-2-L-1-Q). A home-built extracavity EDFA is additionally placed in the recirculation path. The intracavity EDFA compensates for optical loss within the Vernier cavity, whereas the extracavity EDFA compensates for the loss in the recirculation path. The slow-axis-polarized OFC propagates clockwise through the Vernier cavity on the forward pass. After reflection by the partial reflecting mirror (PRM; AFR, PMOFM-55-2-B-Q-F-90), it re-enters the cavity and propagates counterclockwise on the return pass. The same fiber-interferometer-stabilized optical reference used for the fiber OFC is also used to stabilize the Vernier cavity. In contrast to the OFC, the optical reference is polarized along the fast axis of the PMF and propagates clockwise through the Vernier cavity. After circulating through the Vernier cavity, a fraction of the optical reference is extracted through the 99:1 fiber coupler, detected by a photodetector (PD; Thorlabs, PDA10CF-EC), and used to generate the error signal for PDH locking. The optical path length of the Vernier cavity is actively controlled using a PZT actuator through PDH locking. This common optical reference maintains the relative frequency relationship between the OFC modes and the cavity resonances, enabling stable Vernier mode filtering. The ODL provides control of the cavity FSR, which, together with independent control of $f_{rep}$, is used to set the Vernier filtering condition and the resulting filtering interval, as described in the following subsection. After the first clockwise pass through the Vernier cavity, the Vernier-filtered comb is predominantly reflected by the PRM and redirected into the cavity. The reflected comb then propagates counterclockwise through the cavity, undergoing a second Vernier-filtering pass before being extracted as the double-pass Vernier-filtered comb output. This recirculation scheme enhances the effective Vernier filtering by subjecting the comb to two successive filtering passes.

Figure 3(c) shows a representative RF spectrum of the Vernier-filtered comb at an EDFA pump current of 300 mA. The densely spaced modes of the original fiber OFC are converted into a filtered comb with an effective mode spacing of approximately 1 GHz. The intermediate modes are strongly suppressed, demonstrating clear spectral-mode selection through the Vernier effect. Figure 3(d) shows the SMSR as a function of the intracavity EDFA pump current. The SMSR remains nearly constant at pump currents below approximately 70 mA. This behavior is attributed to optical loss in the Vernier cavity dominating over optical amplification in the low-pump-current regime. Above approximately 70 mA, the amplification effect becomes significant, and the SMSR increases markedly with increasing pump current, eventually exceeding 70 dB. These results demonstrate the effectiveness of intracavity EDFA for achieving high-SMSR Vernier mode filtering. Having established high-SMSR mode filtering, we next investigate control of the filtering interval through coordinated adjustment of the cavity FSR and the OFC repetition frequency.

2.4 Tunable Vernier mode filtering

In the present system, the cavity FSR and $f_{rep}$ can be independently controlled using separate actuators, allowing the Vernier filtering interval to be set by appropriately selecting their combination. The cavity FSR is varied within a range of 30.0–30.6 MHz by changing the optical path length using the ODL incorporated in the cavity, as shown in Fig. 3(b), while $f_{rep}$ of the fiber OFC is independently adjusted within a range of 99.7–100.8 MHz. According to Eq. (2), different combinations of the cavity FSR and $f_{rep}$ establish different integer coincidence conditions between the OFC modes and the cavity resonances, thereby providing different Vernier filtering intervals. In the present system, the Vernier filtering interval $f_{Vernier}$ was tunable from approximately 0.3 to 3.5 GHz.

Figure 4 shows representative RF spectra of the Vernier-filtered comb obtained for different combinations of the cavity FSR and $f_{rep}$. Filtering intervals of approximately 0.7, 1.1, 1.9, and 2.5 GHz were obtained, as shown in Figs. 4(a)–4(d), respectively. The corresponding cavity FSR and $f_{rep}$ values are indicated in each panel. These results demonstrate gigahertz-scale control of the Vernier filtering interval through coordinated adjustment of the cavity FSR and the OFC repetition frequency. This tunability allows the effective mode spacing of the filtered comb to be adapted to the mode-selection conditions required for subsequent optical injection locking.

2.5 Dual optical injection locking

Figure 5(a) shows the experimental configuration of the dual optical injection-locking system. The Vernier-filtered comb described in the preceding subsection is used as the master optical signal for two distributed-feedback (DFB) slave lasers, OIL #1 (LD-PD, DWDM-HPDFB-1558.78-A-PA-WOI; wavelength range = 1557–1561 nm, linewidth = 47.6 kHz, output power = 89.6 mW) and OIL #2 (LD-PD, DWDM-HPDFB-1561.22-A-PA-WOI; wavelength range = 1559–1564 nm, linewidth = 12.0 kHz, output power = 76.3 mW).The gigahertz-scale effective mode spacing produced by Vernier filtering strongly suppresses the intermediate comb modes and thereby facilitates single-mode injection locking to a selected comb mode. The injection current and temperature of each slave laser are independently controlled so that OIL #1 and OIL #2 are injection-locked to two different modes of the Vernier-filtered comb. The outputs of the two OIL lasers are combined and spectrally conditioned using a programmable optical filter (WaveShaper, Coherent, 1000A), followed by amplification with an EDFA (Thorlabs, EDFA100P) and attenuation with a variable optical attenuator (ATT). The resulting dual-carrier optical signal is characterized using an optical spectrum analyzer (OSA; Yokogawa, AQ6360). In addition, the OIL output is heterodyned with an external-cavity laser diode (ECLD; Santec, TSL-570, linewidth = 200 kHz), and the resulting beat signal is detected by a photodetector (PD; Thorlabs, RXM10AF) and analyzed using an electrical spectrum analyzer (ESA; Agilent, E4440A). In contrast to the RF spectra in Figs. 3 and 4, which characterize the repetition-frequency components of the Vernier-filtered comb, this heterodyne measurement enables the spectral purity of the OIL output in the optical-frequency domain to be evaluated.

Figure 5(b) shows the optical spectrum of the dual OIL output measured with an OSA at a resolution (RES) of 0.1 nm. Two dominant optical carriers are observed at approximately 1561.2 and 1562.9 nm. These carriers correspond to OIL #1 and OIL #2 injection-locked to two different modes of the Vernier-filtered comb, respectively. The optical-frequency separation between the two carriers is approximately 209 GHz. The generation of these two spectrally separated optical carriers provides the dual optical input required for subsequent difference-frequency generation by photomixing in the UTC-PD.

Figure 5(c) shows the heterodyne beat spectrum between the OIL output and an ECLD. A dominant beat signal corresponding to the selected injection-locked mode is observed with an SNR of approximately 59 dB. Although weaker beat components originating from residual unwanted comb modes are also observed, they are strongly suppressed relative to the desired mode, with a power ratio of approximately 42 dB between the desired beat signal and the strongest unwanted beat component. Because the optical-frequency components are mapped to the RF domain through heterodyne detection with the ECLD, this measurement provides a direct evaluation of the spectral purity of the OIL output in the optical-frequency domain. These results confirm selective single-mode OIL using the Vernier-filtered comb as the master signal. The resulting dual optical carriers with high spectral purity are subsequently used for THz generation by photomixing in the UTC-PD.

## 3. THz Generation and Characterization

### 3.1 THz generation and phase-noise characterization

Figure 6(a) shows the experimental setup for THz generation and characterization. The two optical carriers generated by the dual OIL system described in Section 2.5 are incident on a UTC-PD (UTC-PD; NTT Innovative Devices Corporation, IOD-PMJ-24001, operating frequency range = 220–330 GHz, output power = 0 dBm), where they are photomixed to generate a continuous-wave THz signal at their optical difference frequency. The optical power incident on the UTC-PD is adjusted using an ATT. As a representative frequency, a THz signal at 267 GHz was characterized in detail. The generated THz signal is introduced into a subharmonic mixer (SHM; Virginia Diodes, WR3.4MixAMC-I, frequency range = 220–330 GHz) and down-converted to an intermediate frequency (IF) for spectral and phase-noise characterization. The local-oscillator (LO) signal for the SHM is generated using a microwave synthesizer (Anritsu, MG36021A, frequency range = 0.01–20 GHz; phase noise = –126 dBc/Hz at 10 GHz and 10 kHz offset) followed by a ×24 frequency multiplier (Virginia Diodes, WR3.4MixAMC-I, frequency range = 9.17–13.75 GHz). The down-converted IF signal is analyzed using an ESA and a phase-noise analyzer (PNA; Agilent, E4440A, phase noise floor = –114 dBc/Hz at 1 GHz and 10 kHz offset) for spectral and phase-noise measurements, respectively.

Figure 6(b) shows the IF spectrum obtained by down-converting the 267-GHz THz signal with the SHM. A dominant spectral peak is observed at an IF frequency of approximately 110.4 MHz with a signal-to-noise ratio (SNR) exceeding 70 dB. Several weaker spurious components are also observed, although their origins have not been identified. The high-SNR IF signal demonstrates the generation of a continuous-wave

THz signal through photomixing of the two optical carriers produced by the dual OIL system.

Figure 6(c) shows the SSB phase-noise spectrum of the 267-GHz THz signal, shown by the red curve, over an offset-frequency range from 100 Hz to 1 MHz. The measured phase noise reaches −95.7 dBc/Hz at an offset frequency of 10 kHz. As described in Section 2, the low-phase-noise characteristics of the fiber-interferometer-stabilized optical reference are transferred through the fiber OFC to the selected comb modes and subsequently to the two injection-locked optical carriers. Their optical difference frequency therefore inherits the phase coherence originating from the common optical reference. The measured value of −95.7 dBc/Hz, however, does not necessarily represent the intrinsic phase-noise floor of the generated THz signal because the measurement is also affected by the phase noise of the LO and the measurement system.

To evaluate this measurement limitation, Fig. 6(c) compares the measured THz phase noise (red) with the ×24-multiplied LO phase noise (green), the estimated OFC-derived phase noise including the measured residual phase error (orange), and the ideal optical-reference-derived phase noise (blue), with the latter two referred to 267 GHz. In particular, the ×24 frequency multiplication increases the phase noise of the microwave synthesizer used as the SHM LO by 27.6 dB, as discussed in Section 4.1. The measured THz phase noise approaches the multiplied-LO phase noise, suggesting that the measurement is predominantly limited by the LO system in this frequency region. Consequently, the measured value of −95.7 dBc/Hz should be regarded as a measurement-limited value rather than the intrinsic phase-noise floor of the THz synthesizer, suggesting that the actual THz phase noise may be lower. The expected phase-noise contribution transferred from the optical reference to the approximately 267-GHz optical difference frequency is discussed quantitatively in Section 4.

3.2 Frequency-tunable THz synthesis

The THz carrier frequency can be varied by combining tunable Vernier mode filtering with dual OIL. As demonstrated in Fig. 4, the Vernier filtering condition is controlled by selecting an appropriate combination of the Vernier-cavity FSR and the repetition frequency $f_{rep}$ of the fiber OFC. Different combinations of these two independently controllable parameters provide different effective mode spacings of the Vernier-filtered comb, thereby establishing suitable mode-selection conditions for subsequent OIL. By independently controlling the injection current and temperature of OIL #1 and OIL #2, the two slave lasers can be injection-locked to different selected modes of the Vernier-filtered comb. Consequently, the optical frequencies $\nu_1$ and $\nu_2$ of the two OIL carriers, and hence their frequency separation $|\nu_2 - \nu_1|$, can be selected. Photomixing these two carriers in the UTC-PD generates a THz signal at the corresponding difference frequency. Using this approach, THz generation was experimentally demonstrated at three carrier frequencies of 230, 267, and 320 GHz. The THz signals were characterized using the same measurement system shown in Fig. 6(a).

Figures 7(a)–7(c) show the down-converted IF spectra of the THz signals generated at 230, 267, and 320 GHz, respectively. The corresponding cavity FSR, $f_{rep}$, and $f_{Vernier}$ used for each THz frequency are indicated in the respective panels. A dominant spectral peak is clearly observed at each THz carrier frequency. As described in Section 3.1, the 267-GHz signal exhibits an SNR exceeding 74 dB, while the SNRs of the 230- and 320-GHz signals are 70 and 71 dB, respectively. Several weaker spurious components are also observed, although their origins have not been identified. These results demonstrate high-SNR continuous-wave THz generation at different carrier frequencies by changing the pair of comb modes selected for dual OIL.

Figures 7(d)–7(f) show the corresponding SSB phase-noise spectra at 230, 267, and 320 GHz, respectively, measured over an offset-frequency range from 100 Hz to 1 MHz. At an offset frequency of 10 kHz, the measured phase-noise levels are −96.8, −95.7, and −96.8 dBc/Hz at 230, 267, and 320 GHz, respectively. The three phase-noise spectra exhibit similar overall characteristics, demonstrating low-phase-noise THz generation at the different carrier frequencies investigated. These results demonstrate the distinct roles of the two frequency-selection stages in the proposed architecture. The combination of the cavity FSR and $f_{rep}$ determines the $f_{Vernier}$ and hence the spectral conditions suitable for selective OIL, whereas the pair of modes selected as master signals for dual OIL determines the final optical-frequency separation and THz carrier frequency. Thus, adjustment of the Vernier filtering condition does not directly determine the THz frequency; rather, it provides flexible mode-selection conditions for selecting the desired optical carriers from the dense fiber OFC. The broader frequency-selection capability enabled by the independent control of the cavity FSR and $f_{rep}$ is discussed in Sec. 4.2.

## 4. Discussion

4.1 Phase-noise transfer and noise limitations

The phase-noise performance of the generated THz signal is fundamentally related to the phase-noise characteristics of the optical frequency reference. In the present system, a CW laser at a wavelength of approximately 1550 nm, corresponding to an optical frequency of approximately 193.4 THz, is stabilized to a long-fiber interferometer and serves as the common optical frequency reference for the fiber OFC and the Vernier cavity. As shown in Fig. 2(b), the stabilized optical reference exhibits a SSB phase noise of −83.6 dBc/Hz at an offset frequency of 10 kHz. Assuming ideal phase-noise transfer without additional noise, frequency division from the optical reference frequency of 193.4 THz to the OFC repetition frequency of approximately 100 MHz reduces the phase noise by

$$\Delta L_{FD} = -20log_{10}\left(\frac{193.4\ THz}{100\ MHz}\right) = -125.7\ dB. \tag{3}$$

Subsequent frequency scaling from the repetition frequency to the experimentally demonstrated THz frequency of 267 GHz increases the phase noise by

$$\Delta L_{FM} = +20log_{10}\left(\frac{267\ GHz}{100\ MHz}\right) = +68.5\ dB. \tag{4}$$

Accordingly, the phase-noise contribution originating from the optical reference is expected to be

$$L_{267GHz} = -83.6 - 125.7 + 68.5 = -140.8\ dBc\ /\ Hz \tag{5}$$

at an offset frequency of 10 kHz. The corresponding ideal optical-reference-derived phase-noise spectrum is shown by the blue curve in Fig. 6(c). This calculated spectrum represents the phase-noise level expected under ideal noise-transfer conditions, in which the optical-reference phase noise is transferred through the OFC without additional noise, and therefore does not directly represent the intrinsic phase noise of the actual THz source.

In the actual system, the phase-noise transfer from the stabilized optical reference to the fiber OFC is affected by the finite servo bandwidth of the phase-lock loop, residual phase error, and noise introduced by the actuators and control electronics. Therefore, the ideal optical-reference-derived phase-noise spectrum alone does not indicate how faithfully the phase fluctuations of the optical reference are transferred to the OFC. To estimate the phase-noise transfer in the actual system, the residual phase-noise spectrum of the optical beat signal $f_{beat}$ between the stabilized optical reference and the phase-locked OFC mode was measured. This beat signal represents the residual relative phase fluctuations between the optical reference and the OFC mode and therefore provides a measure of how faithfully the OFC mode follows the optical reference. Assuming that the phase fluctuations of the optical reference and the residual beat are uncorrelated, the phase-noise power spectral density of the phase-locked OFC mode can be estimated as

$$S_{\phi,OFC}(f) = S_{\phi,ref}(f) + S_{\phi,beat}(f), \tag{6}$$

where $S_{\phi,ref}$ and $S_{\phi,beat}$ denote the phase-noise power spectral densities of the stabilized optical reference and the residual optical beat, respectively. The estimated OFC-mode phase noise was then referred to 267 GHz using the net optical-to-THz frequency-scaling factor given by Eqs. (3) and (4). The resulting estimated OFC-derived phase-noise spectrum is shown by the orange curve in Fig. 6(c). At an offset frequency of 10 kHz, the estimated OFC-derived phase noise is approximately −137 dBc/Hz, compared with approximately −140.8 dBc/Hz for the ideal optical-reference-derived phase noise, indicating only a modest degradation due to the residual phase error of the OFC phase lock.

Low-noise phase transfer through optical injection locking has been demonstrated [12,16] and extensively reviewed [17], indicating that the phase coherence of the master signal can be preserved in the injection-locked slave laser under appropriate locking conditions. Low-phase-noise millimeter-wave generation by direct photodetection of an optically stabilized microcomb using a UTC-PD has also been demonstrated at 300 GHz [18]. Furthermore, low-phase-noise THz generation by direct photomixing of two modes from a fiber-referenced Kerr microresonator soliton comb in a UTC-PD has been demonstrated at approximately 560 GHz [19]. These results indicate that low-noise characteristics can be maintained through the OIL and optical-to-THz conversion stages under appropriate operating conditions. Although the heterodyne measurement in Fig. 5(c) confirms selective single-mode OIL with high spectral purity, it does not directly quantify the additive phase noise of the OIL process. Likewise, the additive phase noise of the UTC-PD was not independently characterized in the present experiment. Therefore, although these previous results suggest that large phase-noise degradation is not inevitable in the OIL and photomixing stages, their individual contributions cannot be quantitatively determined in the present experiment.

The measured THz phase noise must therefore be interpreted together with the noise contribution of the measurement system. As shown by the red curve in Fig. 6(c), the measured SSB phase noise of the 267-GHz THz signal is −95.7 dBc/Hz at an offset frequency of 10 kHz, which is substantially higher than both the ideal optical-reference-derived phase noise and the estimated OFC-derived phase noise. However, this difference cannot be directly attributed to phase-noise degradation in the OFC, OIL, or UTC-PD stages because the THz measurement itself is limited by the LO system. In the present THz measurement system, the LO signal for the SHM is generated by ×24 frequency multiplication of a microwave synthesizer. The phase noise of the microwave synthesizer is therefore increased by

$$\Delta L = +20log_{10}(24) = +27.6\ dB. \tag{7}$$

The corresponding ×24-multiplied LO phase-noise spectrum is shown by the green curve in Fig. 6(c). The measured THz phase-noise spectrum approaches this LO-related noise floor, indicating that the measured value of −95.7 dBc/Hz does not directly represent the intrinsic phase-noise floor of the THz synthesizer. Rather, the present measurement is predominantly limited by the LO and the associated measurement system. Although the estimated OFC-derived phase noise is approximately −137 dBc/Hz at a 10-kHz offset, the present LO-limited measurement does not allow us to determine whether the intrinsic THz phase noise approaches this level. Further characterization using a lower-noise THz measurement system will therefore be required to determine the intrinsic phase-noise performance and to separately quantify the additional phase-noise contributions introduced by the OIL and UTC-PD stages.

4.2 Frequency selection and tunability of the Vernier-based THz synthesizer

The dense mode spacing of the fiber OFC provides inherent flexibility in selecting optical-frequency pairs for THz difference-frequency synthesis. With a repetition frequency of approximately 100 MHz, the OFC provides a large number of closely spaced optical modes, allowing different mode pairs to be selected for generating different THz frequency separations. However, this dense mode structure presents a challenge for OIL because multiple OFC modes can fall within the locking range of the slave laser, increasing the possibility of competition among neighboring modes and making selective injection locking to a single desired comb mode difficult. Thus, there is an inherent trade-off between the frequency-selection flexibility provided by the dense OFC spectrum and the spectral selectivity required for stable single-mode OIL. In the present system, Vernier mode filtering bridges these two requirements by converting the densely spaced OFC spectrum into a mode-selective spectrum suitable for subsequent OIL while retaining the underlying frequency-selection capability of the fiber OFC.

As demonstrated in Fig. 3, Vernier mode filtering converts the densely spaced OFC modes with a spacing of approximately 100 MHz into a filtered comb with an expanded effective mode spacing in the gigahertz range, while achieving an SMSR exceeding 70 dB under an appropriate filtering condition. The resulting increase in spectral contrast between the selected mode and the surrounding unwanted modes reduces their contribution within the OIL locking range and thereby facilitates selective single-mode OIL. Importantly, the Vernier cavity functions as a mode-selective prefilter rather than as an independent frequency reference. The frequencies and coherence of the selected modes remain determined by the fiber OFC, whose frequency stability and low-phase-noise characteristics are derived from the common fiber-interferometer-stabilized optical reference. Thus, Vernier filtering improves the spectral conditions required for OIL without replacing the optical-frequency reference or the phase-noise-transfer mechanism discussed in Sec. 4.1.

A key feature of the present architecture is that the Vernier filtering interval is not determined by the cavity FSR alone, but by the relative relationship between the Vernier-cavity FSR and the repetition frequency $f_{rep}$ of the fiber OFC. In the present system, these two frequency parameters can be independently controlled using separate actuators: the cavity FSR is adjusted by changing the cavity optical path length with the ODL, whereas $f_{rep}$ is independently adjusted through the repetition-frequency control of the fiber OFC. Appropriate combinations of the cavity FSR and $f_{rep}$ therefore establish different Vernier coincidence conditions and, consequently, different effective mode spacings of the filtered comb. As experimentally demonstrated in Fig. 4, Vernier filtering intervals of approximately 0.7, 1.1, 1.9, and 2.5 GHz were obtained by selecting different combinations of these two parameters. Thus, the present approach provides a Vernier mode-selection architecture with two independently controllable frequency parameters, rather than relying on adjustment of the cavity FSR alone.

To generalize this relationship beyond the representative conditions demonstrated in Fig. 4, Fig. 8 maps the accessible Vernier coincidence conditions as a function of the Vernier-cavity FSR and the fiber-OFC repetition frequency $f_{rep}$. Each line represents a continuous set of ($f_{rep}$, FSR) combinations satisfying FSR = $(M / N)f_{rep}$ for a given pair of coprime positive integers $M$ and $N$, and the color indicates the corresponding Vernier filtering interval $f_{Vernier} = M \bullet f_{rep} = N \bullet \mathrm{FSR}$. The map shows that multiple gigahertz-scale Vernier filtering intervals can be established within the experimentally accessible ranges of the cavity FSR and $f_{rep}$. This set of accessible Vernier conditions provides flexibility in establishing suitable mode-selection conditions for subsequent OIL.

Importantly, the Vernier filtering interval does not directly determine the synthesized THz frequency. The role of the Vernier filtering stage is to suppress the intermediate modes of the dense OFC and provide appropriate master-mode candidates for selective OIL. Two desired OFC modes are then selected as master signals for the two slave lasers, generating two injection-locked optical carriers whose frequency difference determines the generated THz frequency according to $f_{THz} = |v_2 - v_1|$. Because Fig. 8 shows that multiple Vernier conditions can be established over the accessible parameter space, the Vernier filtering stage provides flexibility in establishing suitable mode-selection conditions for different THz frequencies. The overall accessible THz frequency range is additionally constrained by the optical-frequency tuning ranges of the slave lasers and the photomixing bandwidth of the UTC-PD. In principle, the proposed architecture can therefore utilize these available ranges while maintaining the mode selectivity required for stable single-mode OIL. The THz frequencies demonstrated in Fig. 7 represent selected examples of this frequency-selection mechanism rather than the full accessible frequency range of the synthesizer.

## 5. Conclusion

We demonstrated a frequency-tunable, low-phase-noise THz synthesizer based on a fiber-interferometer-referenced OFC, an EDFA-integrated Vernier cavity, dual optical injection locking, and photomixing in a UTC-PD. The Vernier mode-filtering scheme converted the densely spaced fiber-OFC modes with a repetition frequency of approximately 100 MHz into a filtered comb with gigahertz-scale effective mode spacing, while achieving an SMSR exceeding 70 dB. By independently controlling the Vernier-cavity FSR and the OFC repetition frequency, different Vernier coincidence conditions were established, providing flexible mode-selection conditions for subsequent single-mode OIL. Using two selected comb modes as master signals for dual OIL, CW THz signals were generated at 230, 267, and 320 GHz. At 267 GHz, an SNR exceeding 70 dB and a measured phase noise of −95.7 dBc/Hz at a 10-kHz offset were obtained, although the phase-noise measurement was limited by the LO system. These results demonstrate that Vernier mode filtering can provide a flexible interface between a densely spaced, low-phase-noise fiber OFC and selective dual OIL, enabling frequency-tunable THz synthesis while preserving the low-noise characteristics derived from the common fiber-interferometer-stabilized optical reference.

**Funding.** Ministry of Internal Affairs and Communications of Japan (JPMI240910001).

**Disclosures.** The authors declare no conflict of interest.

**Data availability.** Data underlying the results presented in this paper are not publicly available at this time but may be obtained from the authors upon reasonable request.

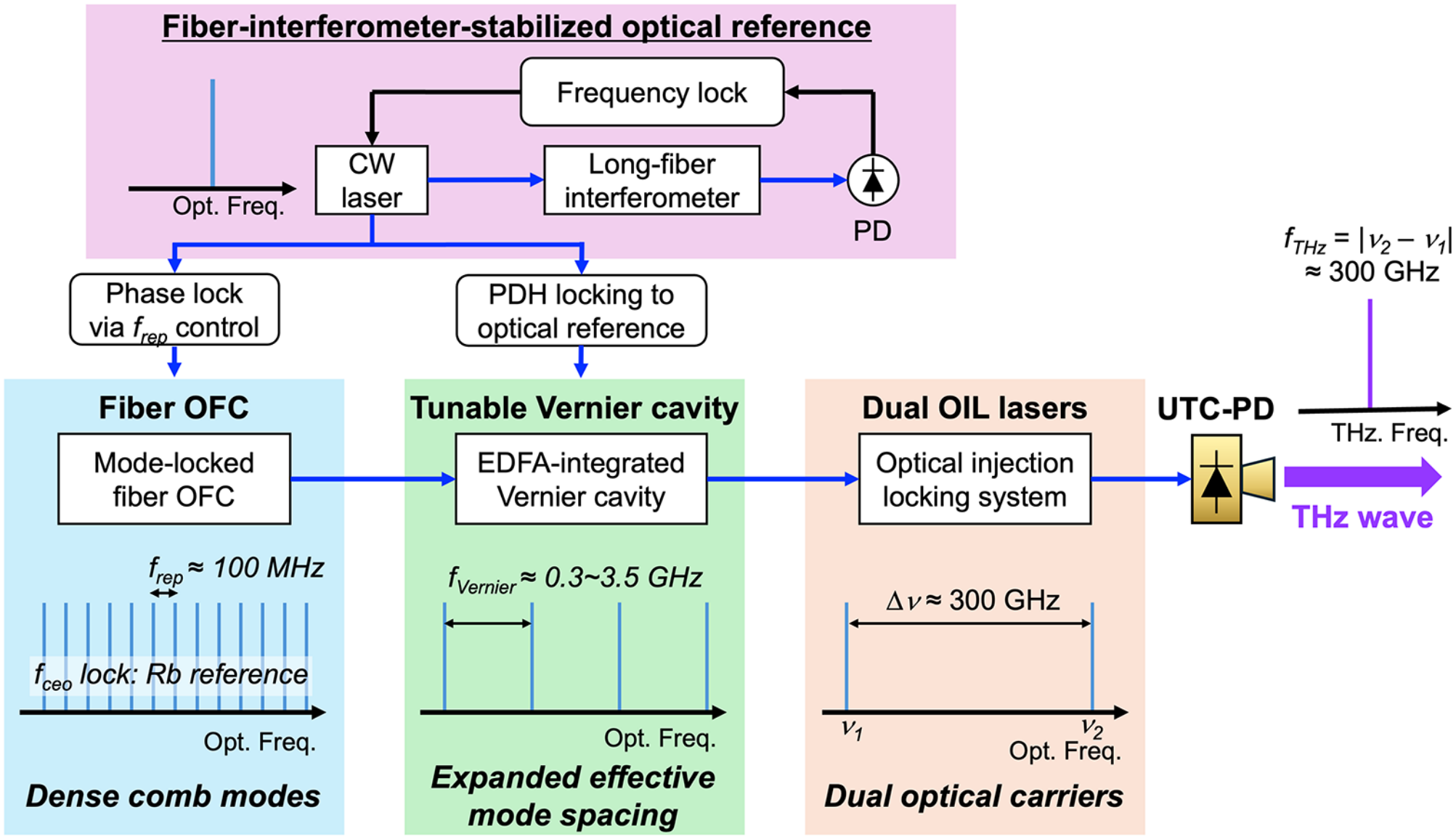

Fig. 1. Conceptual schematic of the tunable low-phase-noise THz synthesizer based on a fiber-interferometer-stabilized optical reference, fiber OFC, EDFA-integrated tunable Vernier cavity, dual optical injection locking (OIL), and UTC-PD photomixing. OFC, optical frequency comb; EDFA, erbium-doped fiber amplifier; UTC-PD, uni-traveling-carrier photodiode; PDH, Pound–Drever–Hall; PD, photodetector; Rb, rubidium.

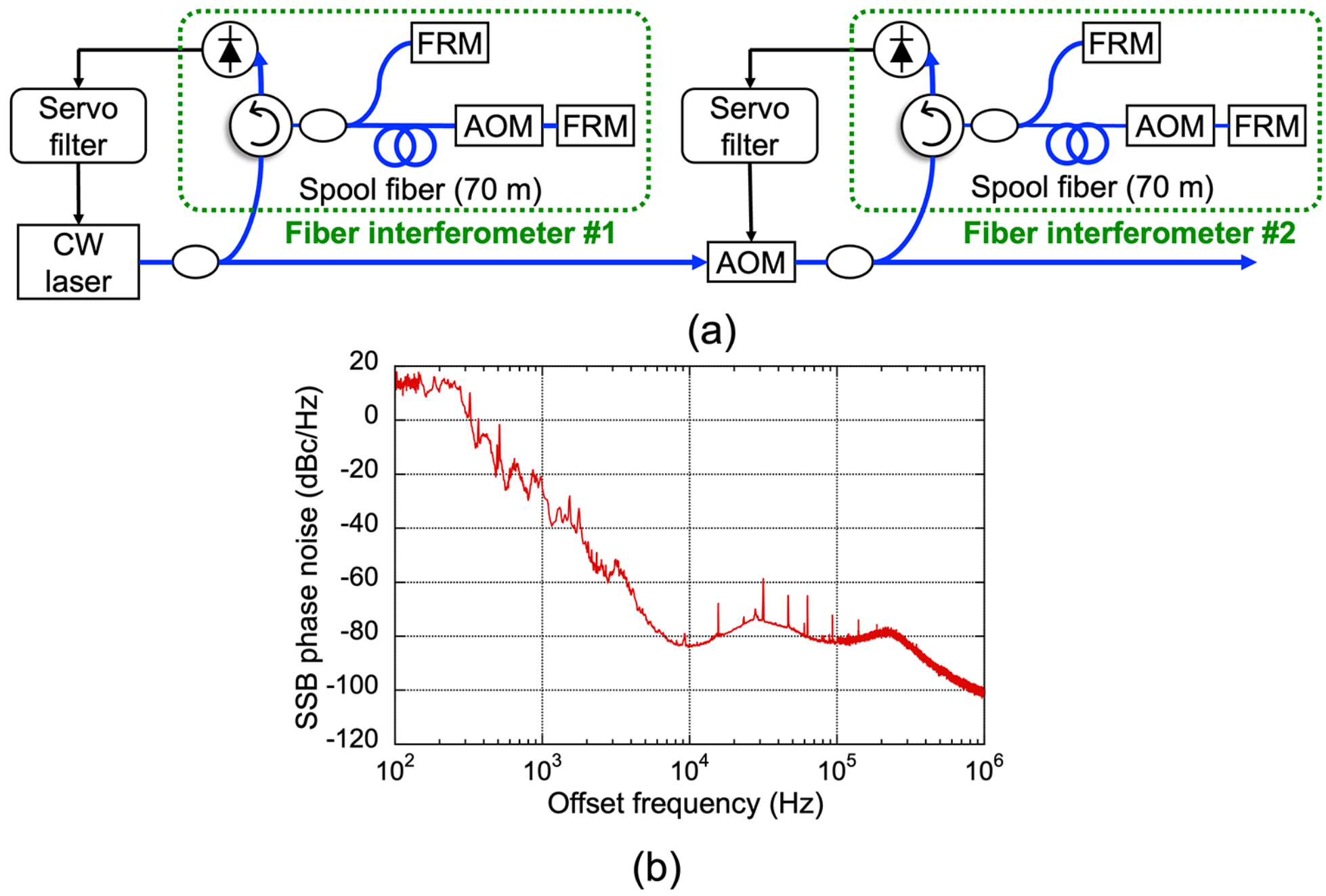

Fig. 2. Fiber-interferometer-stabilized optical reference. (a) Experimental setup of the cascaded two-stage stabilization system using two highly unbalanced fiber Michelson interferometers. (b) SSB phase-noise spectrum of the stabilized CW laser. AOM, acousto-optic modulator; FRM, Faraday rotator mirror; SSB, single-sideband.

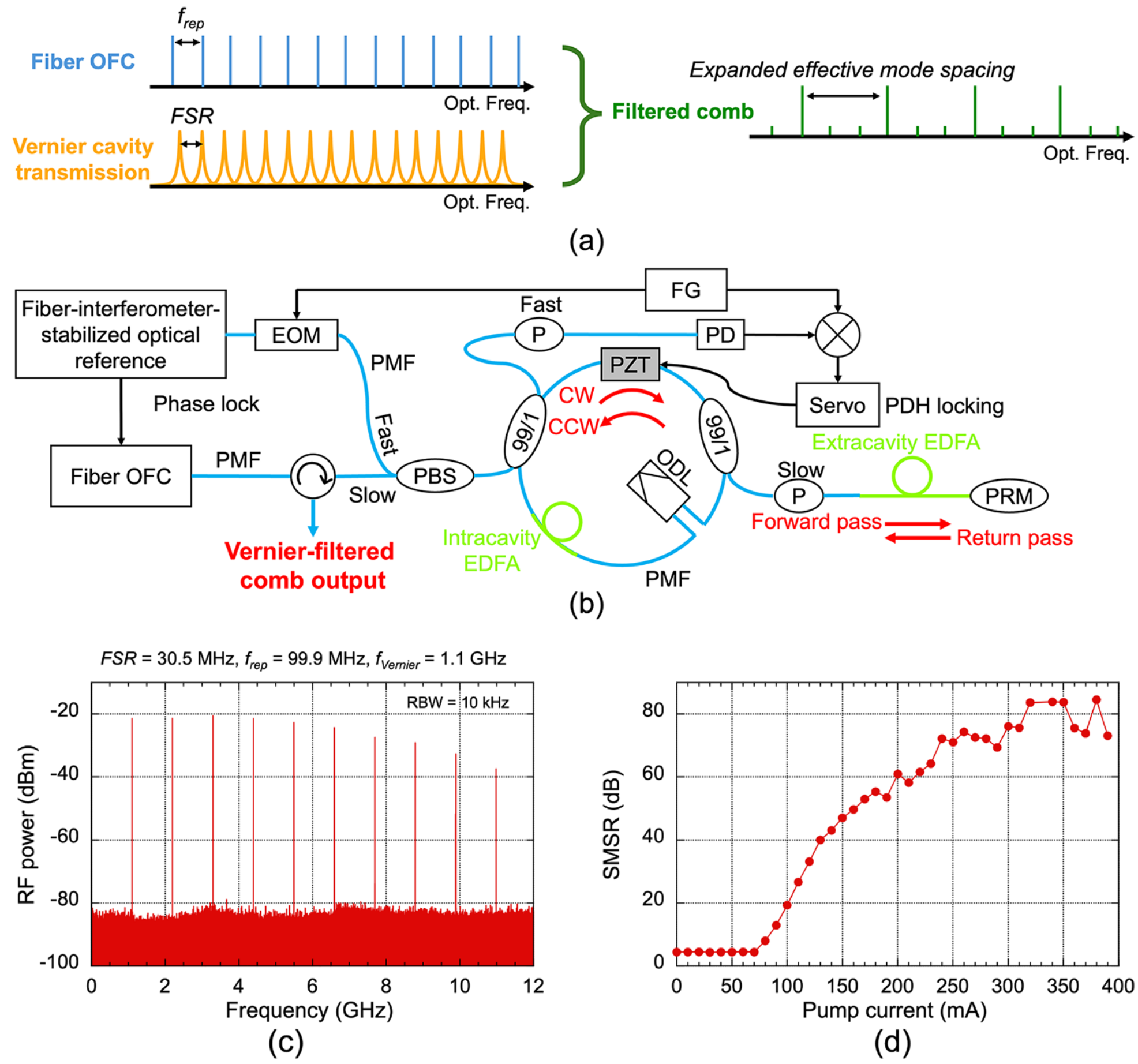

Fig. 3. EDFA-integrated Vernier mode filtering. (a) Principle of Vernier mode filtering and expansion of the effective comb-mode spacing. (b) Experimental setup of the EDFA-integrated Vernier cavity with PDH stabilization. (c) RF spectrum of the Vernier-filtered comb at a Vernier filtering interval of approximately 1.1 GHz. (d) SMSR as a function of the intracavity EDFA pump current. OFC, optical frequency comb; FSR, free spectral range; EDFA, erbium-doped fiber amplifier; EOM, electro-optic modulator; PBS, polarization beam splitter; P, polarizer; ODL, optical delay line; PZT, piezoelectric transducer; PD, photodetector; FG, function generator; PDH, Pound–Drever–Hall; PRM, partial-reflection mirror; RF, radio frequency; RBW, resolution bandwidth; SMSR, side-mode suppression ratio; CW/CCW, clockwise/counterclockwise.

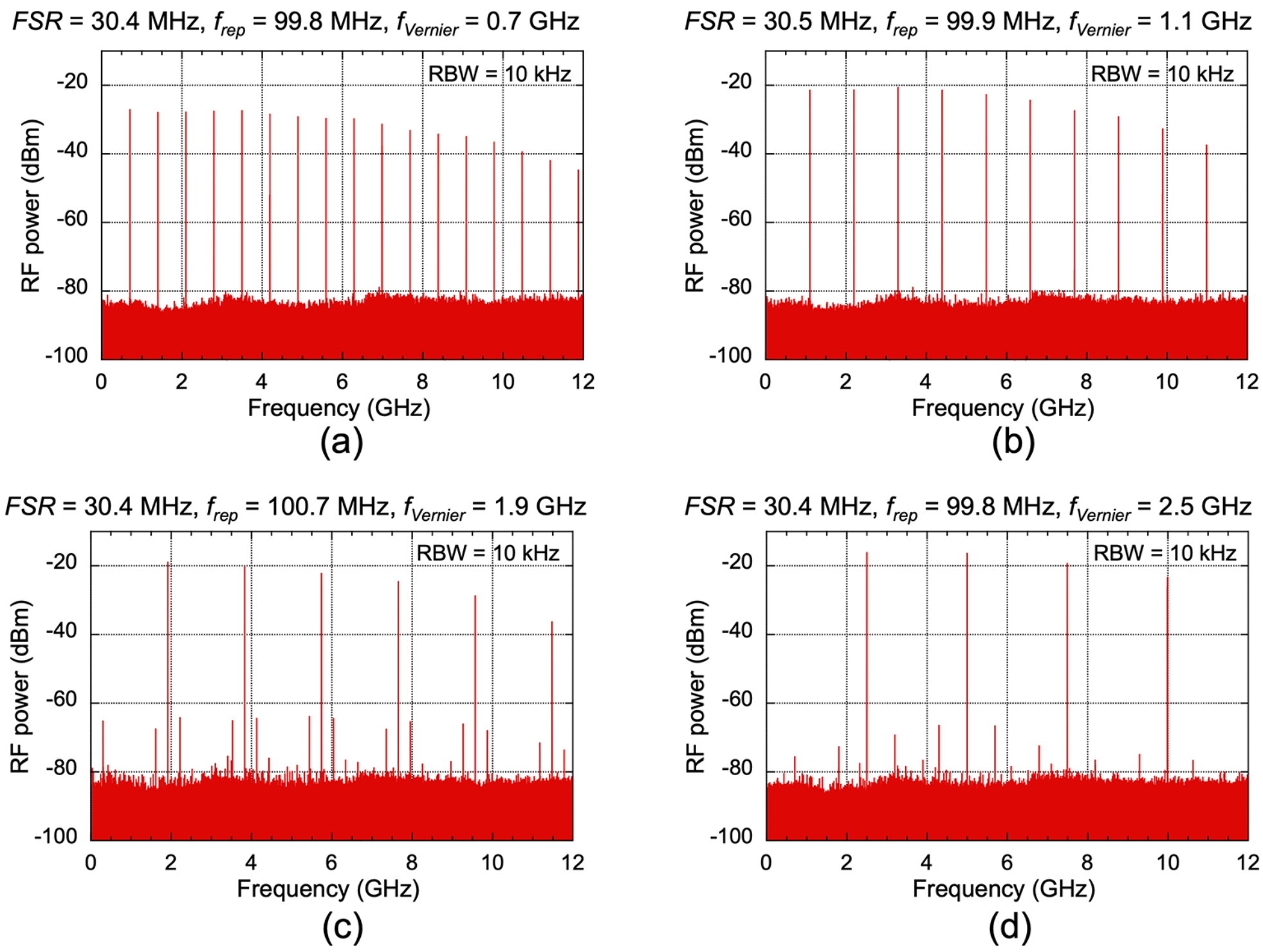

Fig. 4. RF spectra of the Vernier-filtered comb for different Vernier filtering intervals: (a) 0.7 GHz, (b) 1.1 GHz, (c) 1.9 GHz, and (d) 2.5 GHz. FSR, free spectral range; RF, radio frequency; RBW, resolution bandwidth.

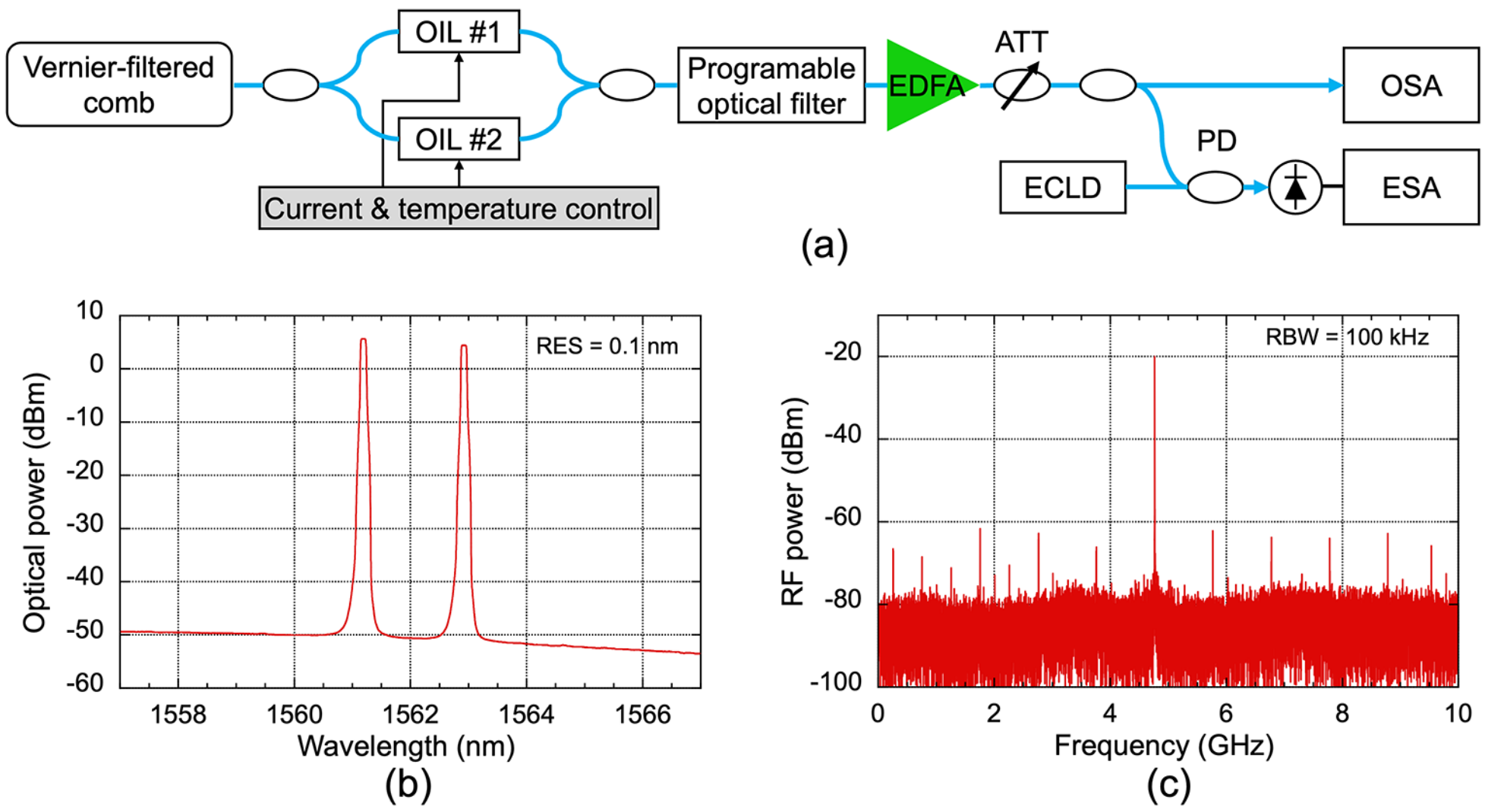

Fig. 5. Dual optical injection locking. (a) Experimental setup for dual OIL. (b) Optical spectrum of the two injection-locked optical carriers. (c) RF spectrum of the heterodyne beat signal between an injection-locked optical carrier and an ECLD. OIL, optical injection locking; EDFA, erbium-doped fiber amplifier; ATT, optical attenuator; OSA, optical spectrum analyzer; ECLD, external-cavity laser diode; ESA, electrical spectrum analyzer; RF, radio frequency.

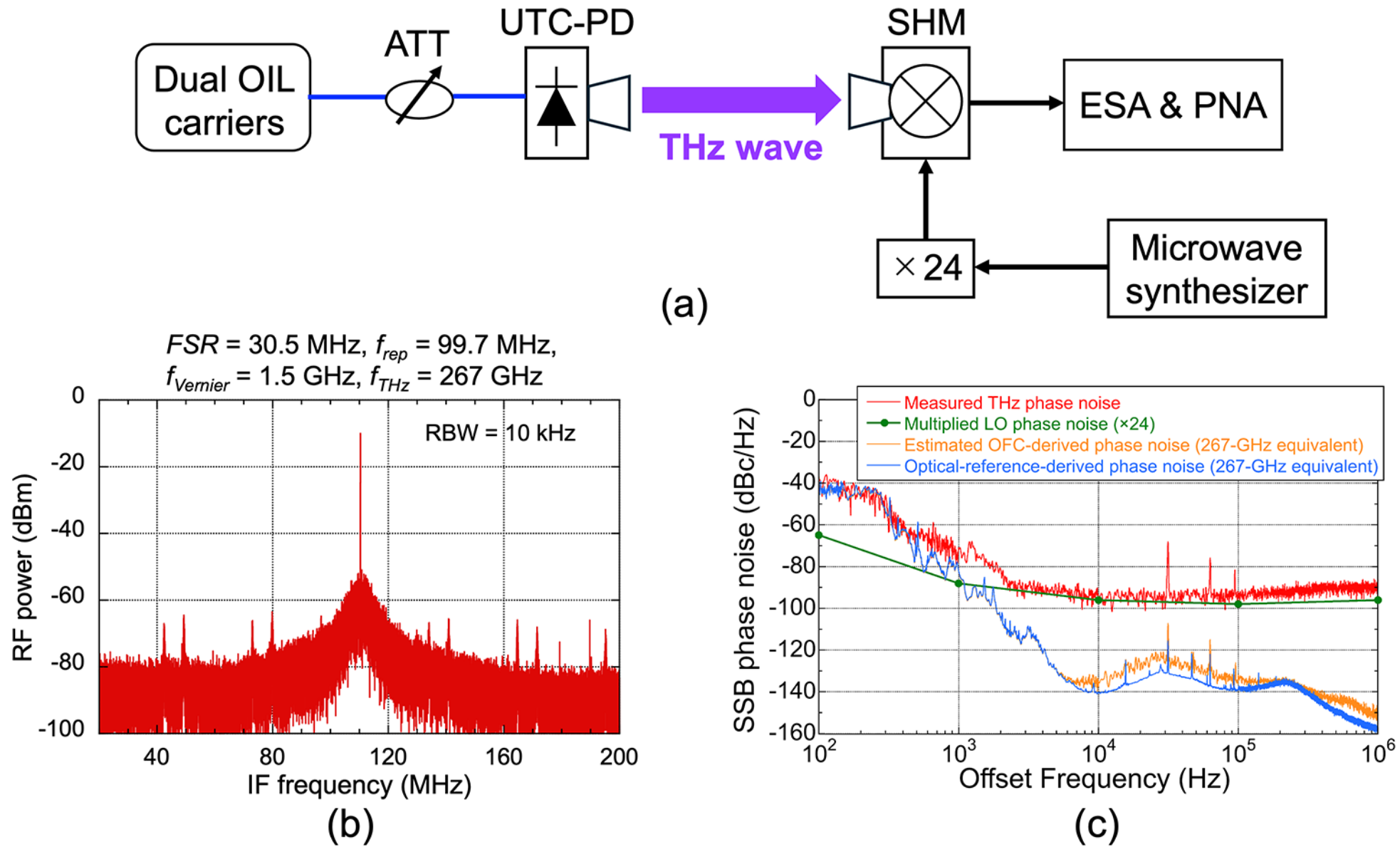

Fig. 6. THz generation and phase-noise characterization. (a) Experimental setup for UTC-PD photomixing and THz heterodyne detection. (b) IF spectrum of the generated 267-GHz THz signal after heterodyne downconversion. (c) SSB phase-noise spectra of the measured 267-GHz THz signal, the ×24-multiplied LO, the estimated OFC-derived contribution including the measured residual phase error, and the ideal optical-reference-derived contribution, with the latter two referred to 267 GHz. OIL, optical injection locking; ATT, optical attenuator; UTC-PD, uni-traveling-carrier photodiode; SHM, subharmonic mixer; ESA, electrical spectrum analyzer; PNA, phase-noise analyzer; LO, local oscillator; IF, intermediate frequency; RBW, resolution bandwidth; SSB, single-sideband.

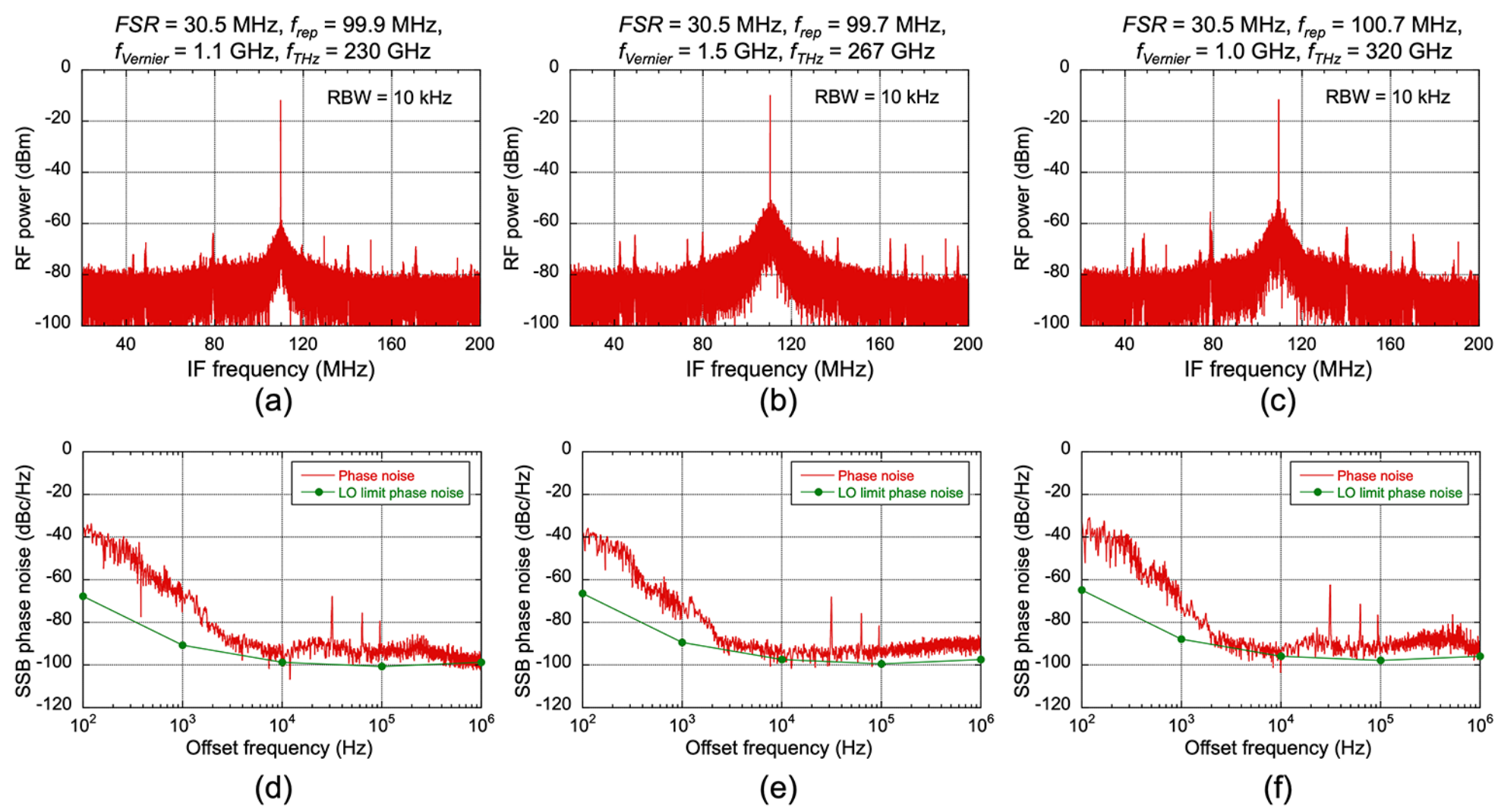

Fig. 7. Frequency-tunable THz synthesis. IF spectra of the generated THz signals after heterodyne downconversion at (a) 230 GHz, (b) 267 GHz, and (c) 320 GHz. Corresponding SSB phase-noise spectra of the measured THz signals and the multiplied LO at (d) 230 GHz, (e) 267 GHz, and (f) 320 GHz. FSR, free spectral range; IF, intermediate frequency; RBW, resolution bandwidth; SSB, single-sideband; LO, local oscillator.

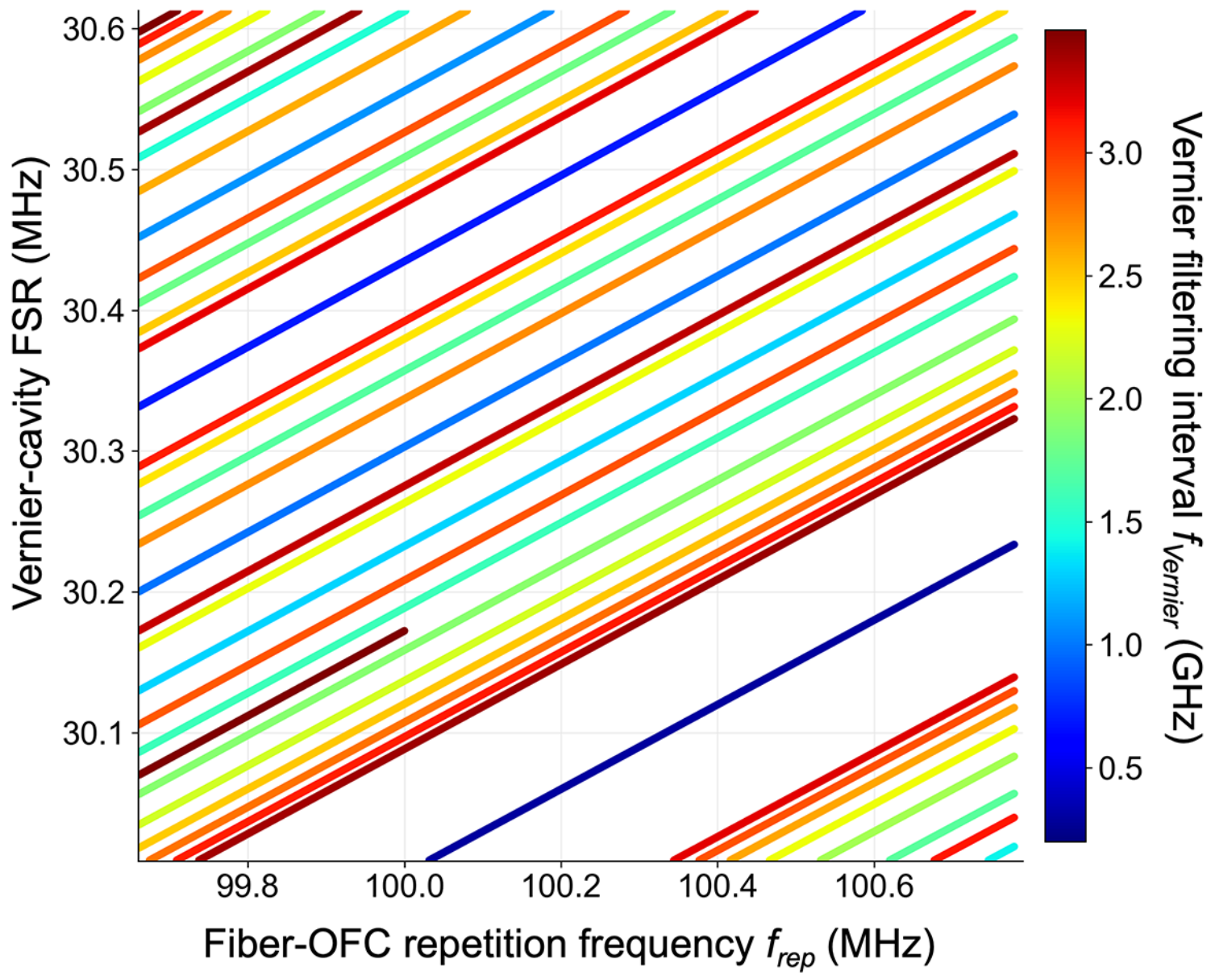

Fig. 8. Accessible Vernier coincidence conditions as a function of the fiber-OFC repetition frequency $f_{rep}$ and Vernier-cavity FSR. Each line represents a continuous set of ($f_{rep}$, FSR) combinations satisfying FSR = ($M$ / $N$)$f_{rep}$ for a given pair of coprime positive integers $M$ and $N$. The color indicates the corresponding Vernier filtering interval $f_{Vernier}$ = $M$•$f_{rep}$ =$N$•FSR. Only Vernier filtering intervals within the range considered in this study are shown.